\documentclass[12pt]{article}
\usepackage{amsmath}
\usepackage{graphicx}
\usepackage{times}
\usepackage[T1]{fontenc}
\usepackage[margin=1cm]{caption}
\usepackage{graphicx}
\usepackage[square,numbers,sort&compress]{natbib}
\usepackage[utf8]{inputenc}
\usepackage{amssymb}
\author
{Kacho Imtiyaz Ali Khan$^{1, \ddag}$, Nidhi Kandwal$^{1}$, Pankhuri Gupta$^{1}$, Deeksha Khandelwal$^{1,2}$,\\ Akash Kumar$^{3,4}$,  Johan \AA kerman$^{3,4}$, and Pranaba Kishor Muduli$^{1\ast}$
\\
\normalsize{$^{1}$Department of Physics, Indian Institute of Technology Delhi, Hauz Khas,} \\\normalsize{ 110016, New Delhi, India}
\\
\normalsize{$^{2}$Inter-University Accelerator Center, New Delhi 110067, India}
\\
\normalsize{$^{3}$Department of Physics, University of Gothenburg, Fysikgr\"and 3, 412 96,} \\\normalsize{Gothenburg, Sweden}
\\
\normalsize{$^{4}$Research Institute of Electrical Communication, Tohoku University, 2-1-1 Katahira,}\\
\normalsize{Aoba-ku, Sendai 980-8577 Japan} \\
\normalsize{$^\ddag$Present address: Paul-Drude-Institut f\"{u}r Festk\"{o}rperelektronik, Leibniz-Institut}\\
\normalsize{im Forschungsverbund Berlin e.V., 10117 Berlin, Germany}\\
\normalsize{$^\ast$muduli@physics.iitd.ac.in}
}

\date{}

\begin{document} 
\title{\LARGE\bfseries{Topologically Driven Giant Effective Spin Mixing Conductance in Antiferromagnetic FeSn/Py Heterostructures}} 
\maketitle
\begin{abstract}
The topological semimetal FeSn antiferromagnet, characterized by its kagome lattice, two-dimensional flat bands, and Dirac-like surface states, holds immense promise for spintronic applications. In this work, for the first time, we investigate the spin pumping behavior in epitaxial-FeSn/Py (Ni$_{80}$Fe$_{20}$) heterostructures. We report a giant effective spin mixing conductance (g$^{\uparrow \downarrow}_{\mathrm{eff}}$) of $(116\pm 7)$~nm$^{-2}$, which is nearly one order of magnitude higher than that of standard Pt/Py heterostructures. The insertion of a 3 nm Al spacer layer results in a two-fold reduction in the effective damping, confirming the interfacial origin of the large g$^{\uparrow\downarrow}_{\mathrm{eff}}$. Consistently, we observe an order-of-magnitude higher inverse spin Hall effect voltage in the FeSn/Py system compared to a reference Pt/Py film stack. We attribute the giant g$^{\uparrow\downarrow}_{\mathrm{eff}}$  to the direct interfacing of the Py layer with the topologically active [001]-kagome surface of epitaxial-FeSn. These findings establish the critical role of topologically active interfaces for advanced quantum-material-based spintronic devices.
\end{abstract}

\section*{Introduction}\label{sec1}
\section{Introduction}
Kagome antiferromagnetic (AFM) materials have emerged as a highly versatile platform for realizing massless Dirac fermions, giving rise to nontrivial topological surface states and spin-momentum locking driven by strong spin-orbit coupling.~\cite{wang2024topological, chowdhury2023kagome, cheng2024untangling} Beyond their intriguing electronic structure, antiferromagnets play a crucial functional role in spintronic heterostructures by providing exchange bias to adjacent ferromagnetic layers.~\cite{khadka2020high, nogues1999exchange} This interfacial exchange coupling enables zero-field skyrmion nucleation~\cite{rana2020room} and field-free magnetization reversal,~\cite{van2016field, peng2020exchange} both of which are key requirements for next-generation memory and logic devices. Moreover, AFM material-based devices are predicted to offer high operational speed and low energy consumption due to the absence of net magnetization and the associated suppression of Joule heating.~\cite{kurebayashi2019theory, burkov2016topological, khan2021energy, moore2010birth}

\begin{figure} [t!]
\centering
\includegraphics[width=0.85\columnwidth]{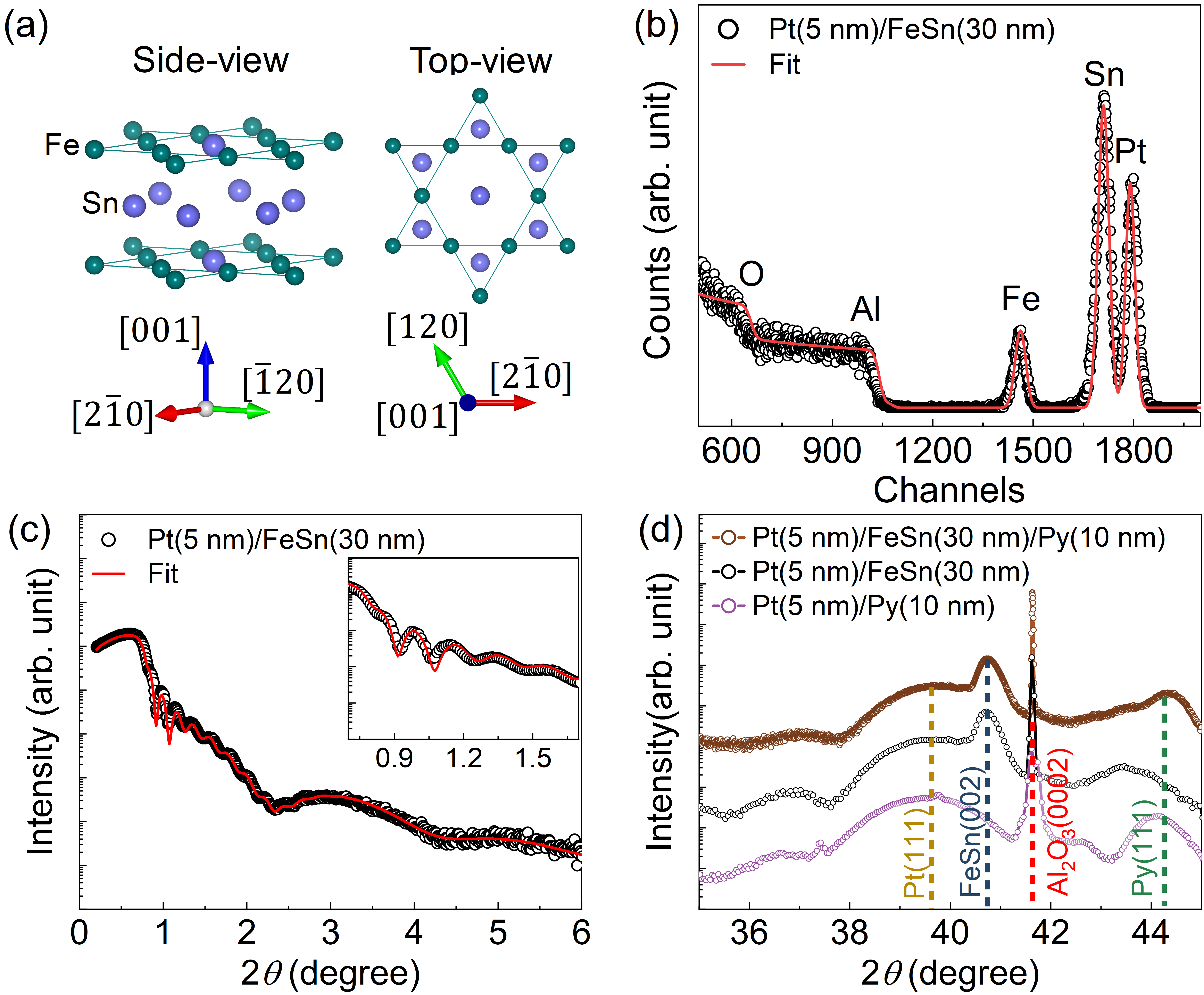}
\caption{(a) Schematic illustration of kagome side-view and top-view of the FeSn crystal structure. (b) Rutherford backscattering spectrometry (RBS) spectrum of the FeSn thin film (open circles) along with the corresponding fit (solid red line). (c) X-ray reflectivity (XRR) spectrum of FeSn film, denoted by circle symbol and fitted with red line. The inset plot shows the zoomed-in view of XRR oscillations corresponds to the FeSn layer. (d) X-Ray Diffraction (XRD) spectra of Pt/FeSn/Py(10 nm), Pt(5 nm)/FeSn(30 nm), and Pt(5 nm)/Py(10 nm) film stack. The short dash vertical lines represent the $2\theta$ positions of Bragg reflections corresponding to the respective planes of Pt, FeSn, and Py layers.}
\label{Fig:1}
\end{figure}

\hspace{1cm}Within the Fe-Sn family of kagome metals,~\cite{zhang2021recent, khan2022intrinsic, agarwal2026spintronic, gupta2025symmetry} FeSn represents a prototypical AFM phase characterized by a collinear magnetic ordering,~\cite{sales2019electronic} in contrast to its sister compound Mn$_3$Sn, which exhibits a non-collinear AFM order.~\cite{bangar2023large} FeSn is a kagome antiferromagnet with a N\'eel temperature of $\sim$368~K and crystallizes into a hexagonal structure (space group $P6/mmm$).~\cite{lin2020dirac} Its crystal structure, depicted in the side-and top-view schematic of Fig.~\ref{Fig:1}(a), consists of Fe$_3$Sn kagome planes that are antiferromagnetically coupled via Sn$_2$ stanene layers. The top-view reveals the characteristic kagome arrangement of Fe atoms within the (001) plane. This specific lattice and magnetic configuration are responsible for generating three-dimensional Dirac fermions in the bulk and topologically protected surface states.~\cite{kang2020dirac, lin2020dirac, tao2023investigating} Beyond its Dirac cone dispersion, the kagome lattice in FeSn gives rise to flat electronic bands, which are direct signatures of enhanced electron correlations.~\cite{Ghimire2020, kang2020dirac, tao2023investigating, lin2018flatbands, do2022damped} Angle-resolved photoemission spectroscopy~\cite{lin2020dirac} and inelastic neutron scattering~\cite{do2022damped} techniques further confirmed their existence and topological origin of electronic flat bands. 
Recent studies have demonstrated that such topological semimetals can strongly enhance spin transport across magnetic interfaces at ferromagnets.~\cite{hong2020large, ding2021large, gupta2025symmetry} For example, a large spin mixing conductance was experimentally observed for Weyl semimetal Mn$_3$Ge/Py heterostructures,~\cite{hong2020large} and similarly, a significant enhancement of Gilbert damping was also observed in Dirac semimetal (DSM) $\alpha$-Sn/Py systems, originating from efficient spin pumping via topological surface states.~\cite{ding2021large} These results suggest that FeSn, as a kagome DSM, is also a promising candidate for realizing an efficient spin sink when interfaced with a conventional ferromagnet, like Py(Ni$_{80}$Fe$_{20}$).~\cite{lin2020dirac, lin2018flatbands} In our previous work, using spin-torque ferromagnetic resonance, we demonstrated a six-fold symmetric damping-like torque and a substantial field-like torque arising from out-of-plane spin polarization, originating from a nodal line with large Berry curvature near the Fermi surface of FeSn.~\cite{gupta2025symmetry} Despite the promising topological band structure properties of the FeSn kagome antiferromagnet, the spin-dynamic properties of epitaxial-FeSn/Py interfaces remain experimentally unexplored and require systematic investigation to fully realize their potential for future quantum-material-based spintronic applications.

\hspace{1cm}In this work, we report on giant spin pumping and enhanced spin-to-charge conversion in antiferromagnetic FeSn/Py heterostructures. High-quality FeSn epitaxial thin films were grown on Al$_2$O$_3$(0001) substrates using a Pt seed layer. Through systematic ferromagnetic resonance (FMR) measurements on FeSn/Py heterostructures, we extract a giant effective spin mixing conductance of g$^{\uparrow\downarrow}_{\rm eff} \approx {(116\pm7)~\rm nm^{-2}}$, nearly an order of magnitude larger than typical heavy-metal/ferromagnet systems. Inverse spin Hall effect (ISHE) measurements demonstrate that spin-to-charge conversion at the FeSn/Py interface correspondingly exceeds that of a reference Pt/Py stack by almost an order of magnitude, driven primarily by the enhanced g$^{\uparrow\downarrow}_{\mathrm{eff}}$. Introducing an Al spacer layer significantly reduces the damping constant, confirming the interfacial origin of the spin pumping, which we attribute to the topologically active FeSn kagome surface states. These findings establish FeSn as a highly efficient spin sink, offering a robust platform for next-generation topological spintronic devices.

\section{Experimental}
High-quality epitaxial FeSn(30 nm) and FeSn(30 nm)/Py($t$~nm) heterostructures were grown on $c$-plane Al$_2$O$_3$(0001) substrates by magnetron sputtering under the optimized conditions also reported previously.~\cite{gupta2025symmetry} The FeSn layer was grown by co-sputtering Fe and Sn targets at a substrate temperature of 550$^\circ$C. To minimize the lattice mismatch between the FeSn layer and the Al$_2$O$_3$ substrate, we always deposited a 5 nm Pt seed layer before FeSn layer. This Pt layer was grown at a substrate temperature of 400$^\circ$C, with a growth rate of 0.27~${\rm \mathring{\mathrm{A}}~s^{-1}}$ (using 60 W of DC power). The ferromagnetic Py layer ($t= 5-15$ nm) was deposited on top of FeSn layer at room temperature with a growth rate of 0.25~${\rm \mathring{\mathrm{A}}~s^{-1}}$ using 40 W DC power. A thickness dependent reference samples of Py($t$~nm) films were grown at room temperature directly on Al$_2$O$_3$(0001) substrates. To directly compare the spin pumping properties at the FeSn/Py interface with those of a standard conventional heavy metal/ferromagnet interface, we also prepared another reference Pt(5 nm)/Py(10 nm) stack grown on an Al$_2$O$_3$(0001) substrate, with similar growth conditions. All films were capped with a 3~nm Al layer deposited at a growth rate of 0.20~${\rm \mathring{\mathrm{A}}~s^{-1}}$ using an RF power of 60 W to prevent oxidation. The base pressure of the sputtering chamber was better than $5 \times 10^{-8}$ Torr, and the working pressure was maintained at $5 \times 10^{-3}$ Torr. During deposition of all films, the substrates were rotated at 60 rpm to ensure a uniform film surface. Rutherford Backscattering (RBS) technique was used to determine the thickness and stoichiometry of FeSn film, employing 2~MeV He$^{+}$ ions. The film thickness was further confirmed using the X-ray reflectivity (XRR) technique. The epitaxial quality of the films was characterized by $\theta$-2$\theta$ XRD measurements using a PANalytical X'Pert diffractometer with Cu $K_\alpha$ radiation ($\lambda = 1.5418$~\AA). The spin-dynamic properties of these film stacks were investigated using a coplanar waveguide (CPW)-based FMR setup.~\cite{akash2019PhysicaB, khan2024magnetodynamic} Inverse spin Hall effect (ISHE) measurements were carried out using electrical contacts formed by copper pads in an inverted sample geometry. A field-modulation method was employed, in which the static magnetic field was modulated by a small AC field generated using a pair of Helmholtz coils.~\cite{bangar2023large, kumar2018large}

\begin{figure*} [t!]
\centering
\includegraphics[width=0.9\columnwidth]{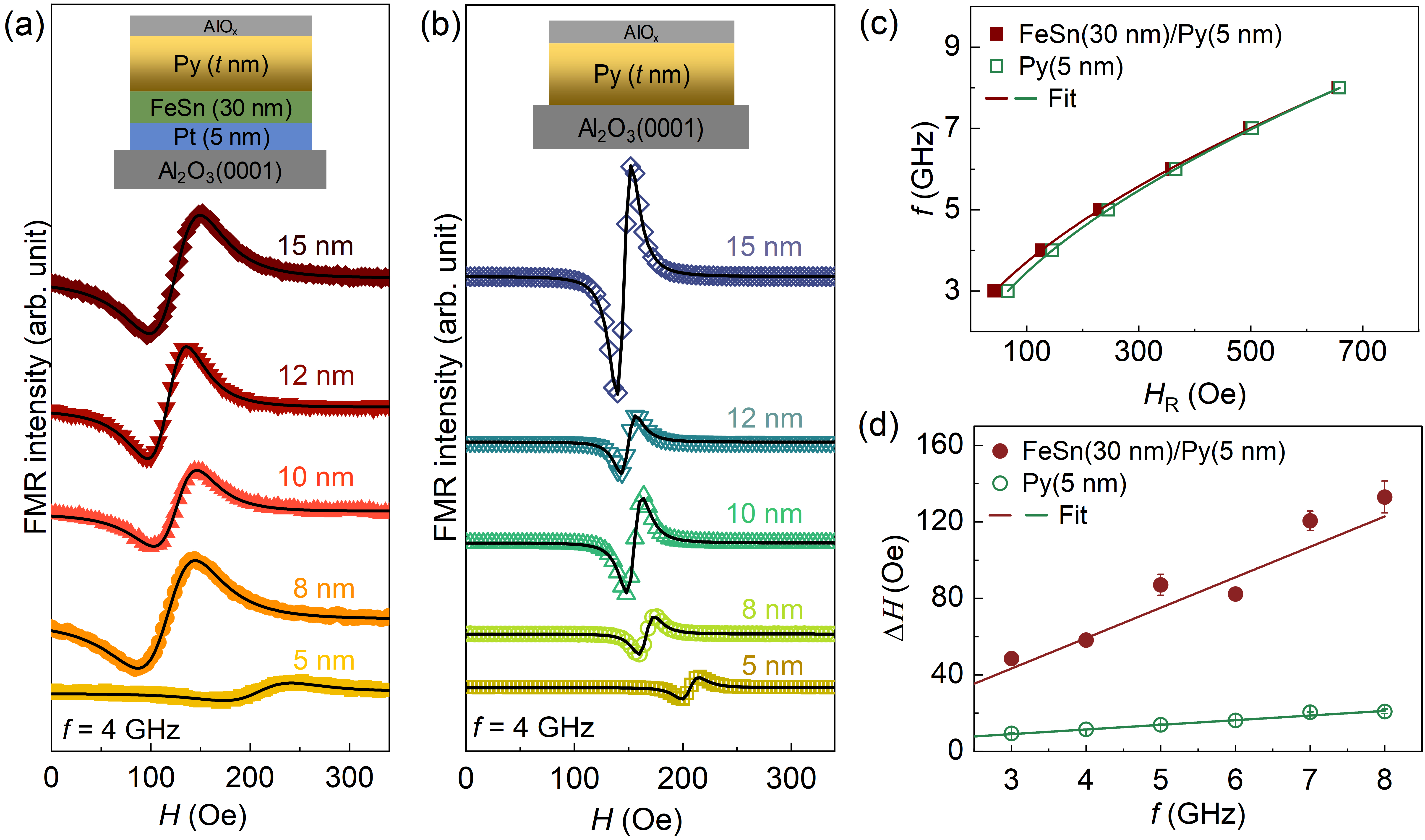}
\caption{\label{fig2} Thickness-dependent ferromagnetic resonance (FMR) spectra measured at $f = 4$~GHz for (a) FeSn/Py($t$~nm) and (b) Py($t$~nm) films, where $t = 15, 12, 10, 8, 5$~nm. The insets show the schematic layer structures of the respective samples. (c) Dependence of the resonance frequency $f$ on the resonance field $H_{\mathrm{R}}$ for FeSn/Py(5~nm) heterostructure, along with the corresponding Py(5~nm) film. (d) Frequency dependence of the linewidth $\Delta H$ for the same samples. The symbols represent the experimental data, while the solid lines denote the fits.
}
\end{figure*}
\section{Results and Discussion}
Figure~\ref{Fig:1}(b) shows the RBS spectrum of the Pt/FeSn film stack, exhibiting well-resolved backscattering edges corresponding to the Pt seed layer, while the Sn and Fe peaks correspond to the FeSn epitaxial layer. The lower-channel signals arise from Al and O of the Al${_2}$O${_3}$(0001) substrate. From the fitting of the RBS spectrum using SIMNRA software,~\cite{Mayer1999} the atomic ratio of Fe:Sn was determined to be 52:48~at.$\%$ with an uncertainty of $\pm$1$\%$, consistent with our previous results.~\cite{gupta2025symmetry} Using the fit, the thickness of FeSn was also calculated and found to be around $30\pm 1$~nm, later confirmed by XRR. Figure~\ref{Fig:1}(c) presents the XRR spectrum of the Pt/FeSn stack, where the shorter-period oscillations are attributed to the FeSn layer, while the longer-period oscillations arise from the Pt layer. By fitting the XRR data as per the recursive Parratt-formalism,~\cite{parratt1954surface} the thickness, interfacial roughness, and density of the FeSn layer were determined to be $(29.6\pm 0.8)$~nm, 2.24~nm, and 8.32~g/cc, respectively. Figure~\ref{Fig:1}(d) shows the XRD $\theta$-2$\theta$ pattern of the Pt(5 nm)/FeSn(30 nm)/Py(10), Pt(5)/FeSn(30 nm), and Pt(5 nm)/Py(10 nm) film stack grown on c-plane Al$_2$O$_3$. A thin Pt seed layer was used to promote epitaxial growth of the FeSn layer. The distinct FeSn(002) and Pt(111) reflections appear at $2\theta = 40.7^\circ$ and $39.8^\circ$, respectively, confirming the epitaxial growth of both Pt and FeSn layers. For the Pt/Py stack, a Py (111) peak and a broad Pt(111) peak was observed. An additional reflection at $2\theta = 44.3^\circ$ in FeSn/Py film stack, corresponding to Py(111) plane, confirms the crystalline phase formation of the Py layer. To confirm the AFM nature of these FeSn layers, we have performed SQUID measurements for the FeSn film without Py layer, see our previous work for more details.~\cite{gupta2025symmetry}

\hspace{1cm}FMR measurements were performed at room temperature by varying the microwave radio frequency ($f$) from 3 GHz to 8 GHz at an interval of 1~GHz.~\cite{khan2024comparative,khan2024magnetodynamic,akash2019PhysicaB} Here, We performed a detailed study of spin-dynamic measurements on FeSn/Py sample stacks with five different Py layer thicknesses to investigate the spin-pumping properties at the interface between the kagome AFM FeSn epitaxial layer and the Py layer. For this study, we used three thickness-dependent sample series: (i) FeSn/Py: Subs./Pt(5~nm)/FeSn(30 nm)/Py($t$), (ii) Ref-Py: Subs./Py($t$), and (iii) FeSn/Al/Py: Subs./Pt(5~nm)/FeSn(30 nm)/Al($t_{\rm Al}$)/Py(10~nm). Here, $t = 15, 12, 10, 8,$ and $5$~nm, and $t_{\rm Al} = 0.5, 1, 2, 3,$ and $5$~nm.

Figure~\ref{fig2}(a) and (b) show the FMR spectra for all thicknesses in FeSn/Py and ref-Py films measured at 4~GHz [insets represent the schematic of the film stack], respectively. These spectra were fitted with the derivative of the Lorentzian function, which consists of the symmetric and antisymmetric parts.~~\cite{kumar2018large, khan2024magnetodynamic, khan2024comparative} From the fitting, we extracted the resonance field ($H_{\rm R}$) and linewidth ($\Delta H$) as a function of $f$. 

\hspace{1cm}The dependence of $H_{\rm R}$ on $f$ for FeSn/Py (5 nm) and Py (5 nm) films is shown in Fig.~\ref{fig2}(c), while results for other thicknesses are presented in Fig.~S1(a-d) in Supplementary Information, along with fits using in-plane Kittel equation~\cite{kittel1949gyromagnetic}.

\begin{equation}\label{eq:Kittel_in}
 f=\frac{\gamma}{2\pi}\sqrt{(H_{\rm R}+H_{\rm K})(H_{\rm R}+H_{\rm K}+4\pi M_{\rm eff})},
\end{equation}
Here, $\gamma$ is the gyromagnetic ratio, while $H_{\rm R}$, $H_{\rm K}$, and $M_{\mathrm{eff}}$ represent the resonance field, anisotropy field, and effective magnetization of the ferromagnet, respectively. For FeSn/Py($15-5$ nm) thickness series, the extracted values of $M_{\rm eff}$ and $H_{\rm K}$ were found to be in the range of $(747-467)$~emu/cc and $(65-89)$~Oe, respectively. While, for Py($15-5$~nm) thickness series, the value was obtained to be in the range of $(783-587)$~emu/cc and $(37-35)$~Oe. A drastic reduction of $M_{\rm eff}$ down to 467~emu/cc for FeSn/Py(5 nm) is due to the large value of $H_{\rm K}$ and could be due to the strong spin-orbit coupling at the interface of FeSn/Py compared to Py($t$~nm) films, similar to reported work on the Dirac semimetal RhSi/Py film stack.~\cite{panda2025spin}  

\hspace{1cm}In Fig.~\ref{fig2}(d), the variation of overall extracted $\Delta H$ with $f$ was fitted with the following equation for FeSn/Py (5 nm) and Py (5 nm) films (see supplementary Fig.~S1(e-h) for other thicknesses),  to extract the effective Gilbert damping constant ($\alpha_{\rm eff}$) given below;\cite{rossing1963resonance,heinrich1985fmr,celinski1991ferromagnetic,mcmichael2003localized}
\begin{equation}\label{eq:FMR_damping}
    \Delta H= \Delta H_{\rm{0}}+\frac{2\pi\alpha_{\rm eff}}{\gamma}f,
\end{equation}
Here, the first term contains the inhomogeneous linewidth ($\Delta H\rm_{0}$) originating from sample inhomogeneities, and the second term is the frequency-dependent part, which yields the value of effective damping $\alpha_{\rm eff}$. Here, in the case of FeSn/Py($t$~nm), the $\alpha_{\rm eff}$ originated from both intrinsic Gilbert damping ($\alpha_{\rm int}$) of FM and spin pumping contribution from FM to AFM. In Series:~2, Py is deposited directly on the substrate, which is expected to have a negligible contribution to spin pumping. The values of $\Delta H\rm_{0}$ were found to be in range of ($2.5-10.2$)~Oe and ($1.6-2.8$)~Oe for both FeSn/Py($t$ nm) and Py($t$ nm), respectively. Furthermore, the values of $\alpha_{\rm eff}$ were determined to be in the range of $[(17.1\pm0.6)-(44.1\pm0.2)]\times10^{-3}$ for FeSn/Py($15-5$ nm) films and [($6.5\pm0.3)-(8.7\pm0.4)]\times10^{-3}$ for Py($15-5$ nm) films. A large increase in $\alpha_{\rm eff}$ for FeSn/Py($t$ nm) films compared to Py($t$ nm) films directly indicates a giant enhancement of the damping constant solely due to the spin pumping effect into the AFM FeSn layer. 
\begin{figure} [b!]
\centering
\includegraphics[width=0.7\columnwidth]{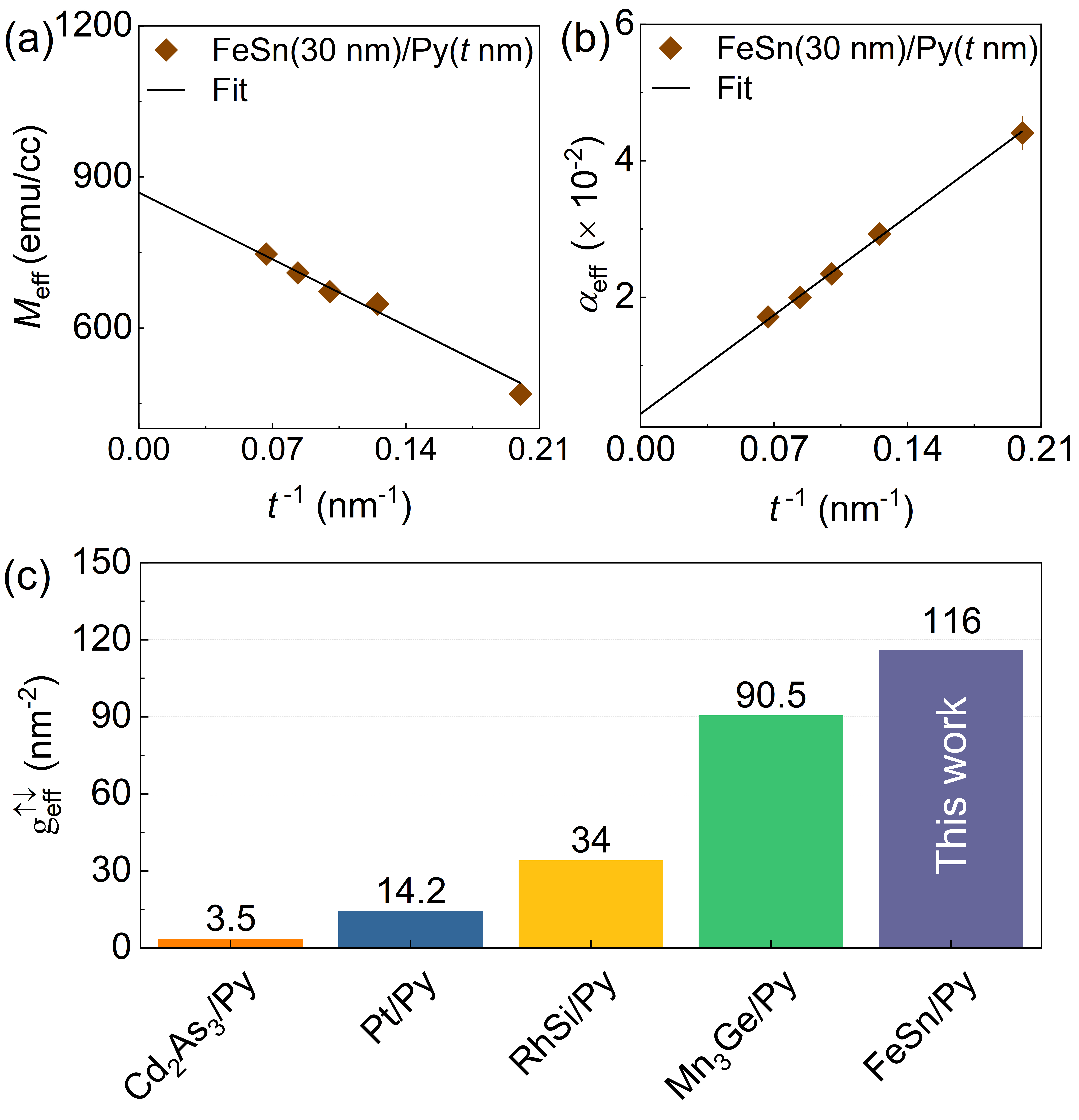} 
\caption{Variation of (a) $M_{\mathrm{eff}}$ and (b) $\alpha_{\mathrm{eff}}$ as a function of the inverse Py thickness ($t$) for FeSn/Py($t$~nm) films. The symbols represent the experimental data, while the solid black lines correspond to the fit using Eq.~\eqref{eq:aniso}\&\eqref{eq:spinmixing}. (c) Comparison of our extracted effective spin-mixing conductance (g$^{\uparrow \downarrow}_{\rm eff}$) with reported values for other Py-based bilayer heterostructures.}
\label{Fig3}
\end{figure}

Figure~\ref{Fig3}(a) shows the extracted values of $M_{\rm eff}$ plotted as a function of $t^{-1}$ for  FeSn/Py($t$~nm) films. The plot was fitted using the expression;
\begin{equation}\label{eq:aniso}
 M_{\rm eff}~=~M_{\rm S}+\frac{2K_{\rm S}}{\mu_{0}M_{\rm S}}\times{t^{-1}},    
\end{equation}
Here, $\mu_{0}$ is the permeability constant of free space. $M_{\rm S}$, and $K_{\rm S}$ are the saturation magnetization and surface anisotropy constant, respectively. For the FeSn/Py film, we found the value of $M_{\rm S}=(869\pm26)$~emu/cc and $K_{\rm S}=(1.03\pm0.01)$~erg/cm$^2$. The extracted $M_{\rm S}$ from FMR measurements are consistent with the value obtained via SQUID magnetometry in our previous work.

To investigate the role of FeSn layer in enhancing the spin pumping contribution at the interface of FeSn/Py bilayer [Figure~\ref{Fig3}(b)], we fit $\alpha_{\rm eff}$ versus $t^{-1}$ using the expression:
\begin{equation}\label{eq:spinmixing}
    \alpha_{\rm eff} = \alpha_{\rm int}+\gamma \hbar \frac{{\rm g}^{\uparrow \downarrow}_{\rm eff}}{4\pi M_{\rm S}}t^{-1},
\end{equation}
The parameter $\hbar$ represents the reduced Planck's constant. Here, we extracted the values of a $\alpha_{\rm int}$ and g$^{\uparrow \downarrow}_{\rm eff}$ from the fitting using Eq.~\eqref{eq:spinmixing}. The value of $\alpha_{\rm int}$ is found to be close to ($2.8\pm0.7$)$\times10^{-3}$. The values of g$^{\rm \uparrow \downarrow}_{\rm eff}$ was found to be ${116\pm7~\rm nm^{-2}}$ for FeSn/Py, which is  almost an order of magnitude higher compared to the previously reported values for HM/Py bilayers: Pd/Py,~\cite{tao2018self,behera2015effect} Pt/Py,~\cite{tao2018self,nakayama2012geometry} and DSM/Py bilayer: RhSi/Py,~\cite{panda2025spin}  and Cd$_3$As$_2$/Py~\cite{yanez2021spin} heterostructures. The value is also larger than previously reported topological Weyl semimetal-based bilayer Mn$_3$Ge/Py.~\cite{hong2020large} The comparitive results are summarized in Fig.~\ref{Fig3}(c).

To make another comparison, we also determined g\(_{\rm eff}^{\uparrow \downarrow}\), using the \(\Delta \alpha = \alpha_{\rm eff} - \alpha_{\rm ref}\) method;~\cite{khan2024comparative} 
\begin{equation}\label{eq:spinmixingalpha}
{{\rm g}^{\uparrow \downarrow}_{\rm eff}}=\frac{4\pi M_{\rm S} t}{\gamma\hslash }(\alpha_{\rm eff}-\alpha_{\rm ref}),
\end{equation}
Here, \(\alpha_{\rm eff}\) is the effective damping of the entire stack [i.e., Pt(5~nm)/FeSn(30~nm)/Py(10~nm) or Pt(5~nm)/Py(10~nm)], while \(\alpha_{\rm ref}\) is the damping contribution only from the ref.-Py layer. For the Pt(5~nm)/FeSn(30~nm)/Py(10~nm) and Pt(5~nm)/Py(10~nm) stacks, we obtained values of \(107.12\pm 0.03~\rm nm^{-2}\) and \(32.06\pm 0.01~\rm nm^{-2}\), respectively. This method also confirms that g\(_{\rm eff}^{\uparrow \downarrow}\) for the FeSn/Py system is almost an order of magnitude higher than that of the standard Pt/Py system.~\cite{tao2018self,nakayama2012geometry}
Therefore, we believe that the origin of giant g$^{\rm \uparrow \downarrow}_{\rm eff}$ in our FeSn/Py bilayer could be a result of topologically active FeSn(001) kagome layer with Dirac points in the momentum space.~\cite{gupta2025symmetry, lin2020dirac} 

\begin{figure} [b!]
\centering
\includegraphics[width=0.7\columnwidth]{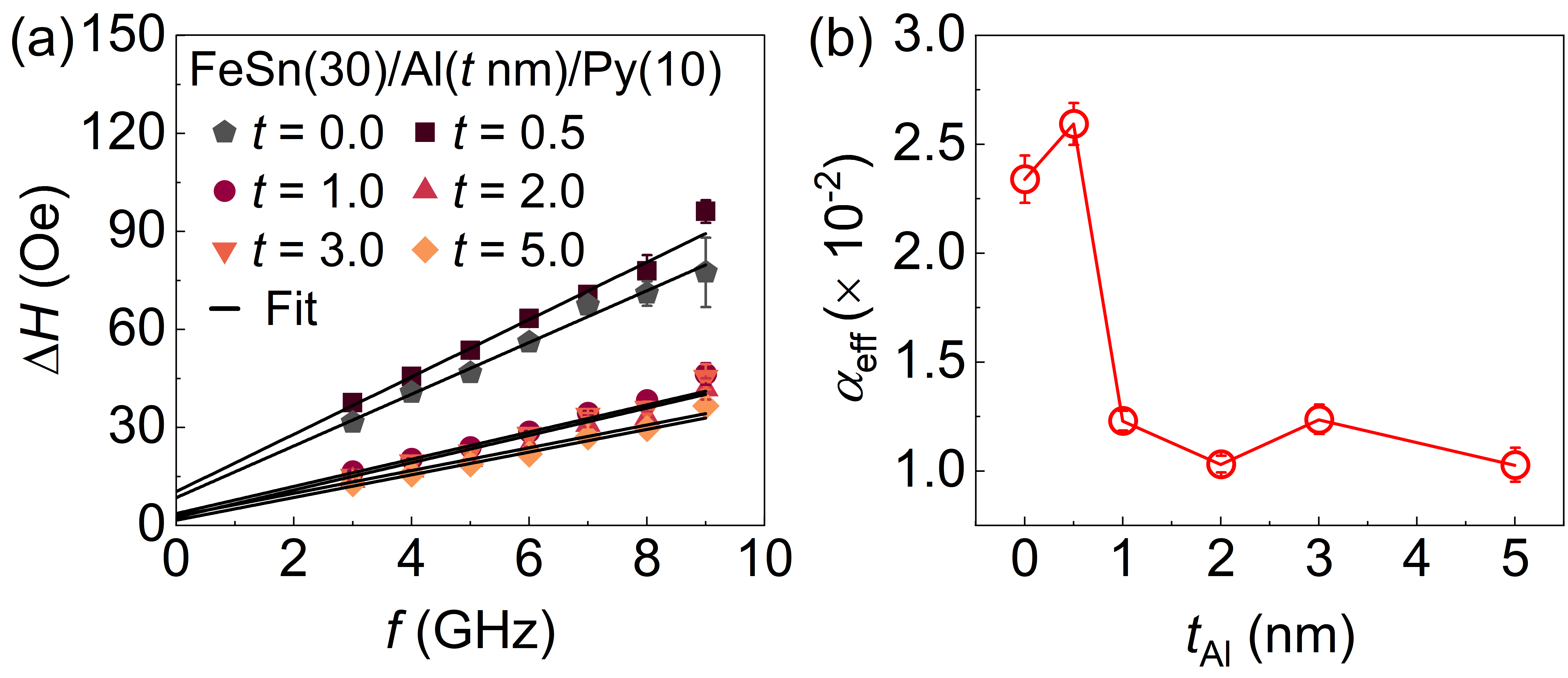}
\caption{
(a) FMR linewidth $\Delta H$ versus \textit{f} for increasing Al spacer for FeSn/Al(\textit{t} nm)/Py(10 nm) trilayer heterostructures; solid lines are fit to the curve using Eqn.~\eqref{eq:FMR_damping}. (b) Extracted $\alpha_{\rm eff}$ as a function of Al spacer thickness.}
\label{Fig4}
\end{figure}
\hspace{1cm}To examine this further, we used trilayer structure FeSn/Al/Py: FeSn/Al($t_{\rm Al}$ nm)/Py(10 nm) with varying thickness of Al spacer and investigated the damping behavior. The obtained values of $\Delta H$ were plotted as a function of \textit{f} and fitted using Eqn.~\eqref{eq:FMR_damping}, as shown in Fig.~\ref{Fig4}(a). A systematic dependence of the extracted $\alpha_{\rm eff}$ as a function of Al spacer thickness is plotted in Fig.~\ref{Fig4}(b). For an ultrathin layer of Al spacer 0.5 nm, a small enhancement in $\alpha_{\rm eff}$ from $(2.34\pm0.11)\times10^{-2}$ to $ (2.59\pm0.09)\times10^{-2}$ was observed, suggesting a small increase in spin pumping contribution at the interface of FeSn/Al(0.5 nm)/Py(10 nm) stack. This enhancement could be attributed to spin-momentum locking occurring due to the topologically active FeSn(001) surface of the FeSn, consistent with prior reports on similar systems, Dirac semi-metal $\alpha-$Sn/Al/Py stacks.~\cite{ding2021large} Moreover, as the Al spacer thickness further increases, $\alpha_{\rm eff}$ decreases sharply to a value of $1.23\times10^{-2}$ and saturates close to this value for Al thicknesses beyond 1 nm. The measured $\alpha_{\rm eff}$ reduced by two-fold when a 3 nm Al spacer was inserted between FeSn and Py, dropping from $1.98\pm0.10$ to $1.03\pm0.06$. This demonstrates that the giant $\alpha_{\rm eff}$ is interfacial in nature, originating from direct coupling of Py to the FeSn kagome surface. 
\begin{figure} [b!]
\centering
\includegraphics[width=0.7
\columnwidth]{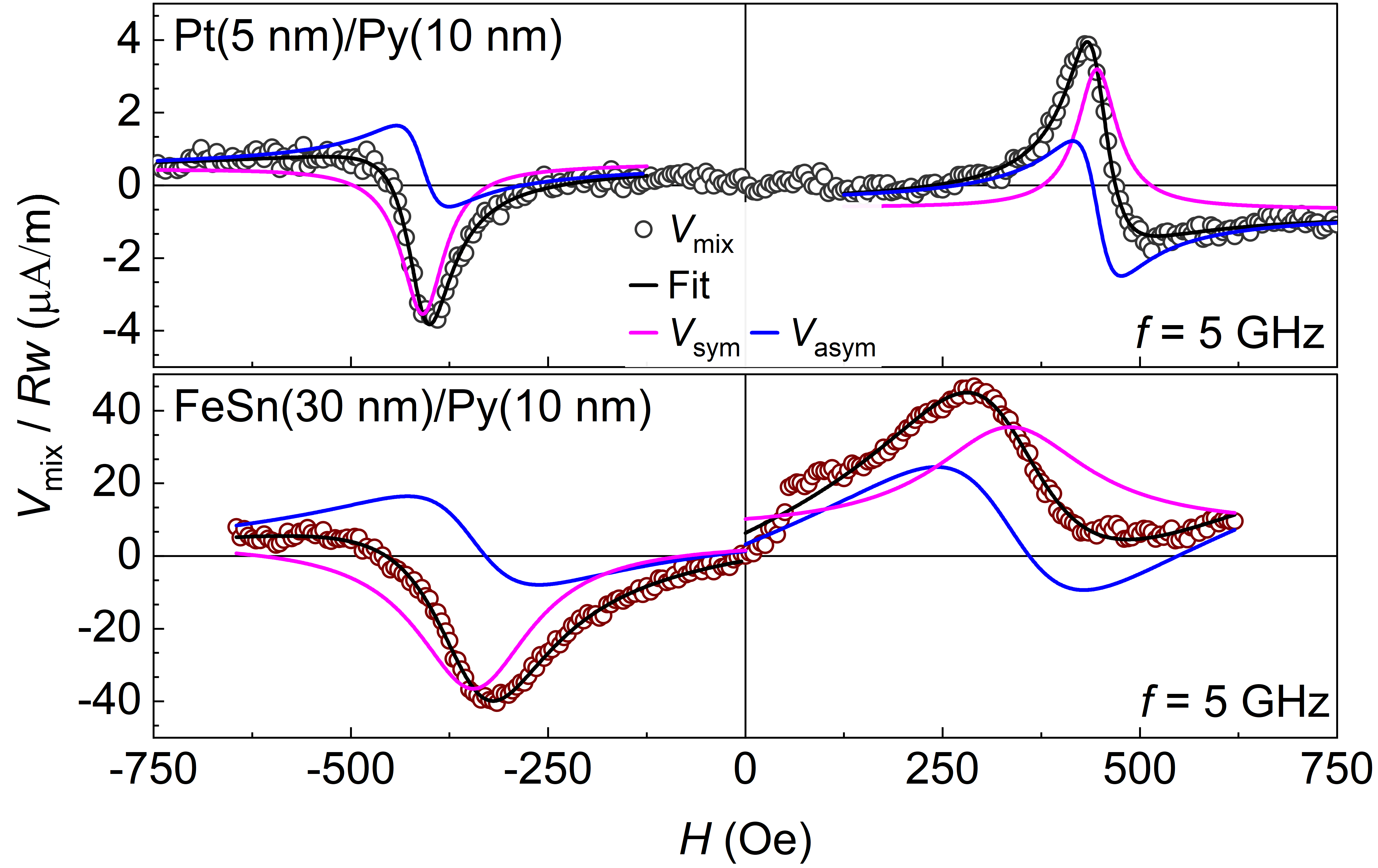} 
\caption{Inverse spin Hall effect (ISHE) spectra measured at $f= 5$~GHz for FeSn(30~nm)/Py(10~nm) and Pt(5 nm)/Py(10~nm) film stacks. The symbols represent the experimental data, while the blue, magenta, and black lines denote the fits corresponds to the antisymmetric-, symmetric-component, and actual spectra using Eq.~\eqref{eq:ISHE}, respectively.}
\label{ISHE}
\end{figure}

\hspace{1cm}To further corroborate the giant spin mixing conductance with efficient spin–charge conversion at the FeSn/Py interface, ISHE measurements were performed.~\cite{saitoh2006conversion} Here, under FMR excitation, a pure spin current $I_{\rm S}$ is generated via spin pumping from the precessing Py layer and injected into the adjacent kagome FeSn layer, where it is converted into a transverse charge voltage 
$V_{\rm mix}$. To eliminate geometrical effects arising from the sample resistance and dimension, the measured voltages were normalized by the sample resistance ($R=20.9~\Omega$) and the sample width ($w=2.53$~nm). Consequently, the quantity $V_{\rm mix}/Rw$, which is proportional to the ISHE-induced charge current, is analyzed in the following. Figure~\ref{ISHE} shows the field-swept ISHE spectra, $V_{\rm mix}/Rw$, measured at $f= 5$~GHz for FeSn(30 nm)/Py(10~nm) and Pt(5~nm)/Py(10~nm) bilayers, where Pt/Py serves as a benchmark system owing to the well-established spin Hall response of Pt. The measured spectra of FeSn/Py(10~nm) and Pt(5~nm)/Py(10~nm) were further fitted using;~\cite{bangar2023large}

\begin{equation}
\label{eq:ISHE}
\frac{V_{\rm mix}}{Rw} =
\frac{V_{\rm asym}}{Rw}
\frac{2\Delta H (H-H_{\rm R})}{(H-H_{\rm R})^2+\Delta H^2}+\frac{V_{\rm sym}}{Rw}
\frac{(\Delta H)^2}{(H-H_{\rm R})^2+\Delta H^2},
\end{equation}
The symmetric component $V_{\rm sym}$ originates from the ISHE, while the antisymmetric part $V_{\rm asym}$ is mainly associated with rectification effects such as the anomalous Hall effect (AHE)~\cite{bangar2023large, kumar2018large}. As seen in Fig.~\ref{ISHE}, the FeSn/Py bilayer exhibits a substantially larger symmetric component compared to the Pt/Py sample, directly indicating an enhanced ISHE voltage, which is expected due to direct contact of the Py layer with the FeSn kagome layer. The normalized ISHE-induced charge current density was extracted from the symmetric component using: $I_{\rm C}= V_{\rm sym}/Rw$, where $V_{\rm sym} = [V_{\rm sym}(+H)+V_{\rm sym}(-H)]/2$. Using these values, we determined the normalized value $I_{\rm C}=34.3\pm 0.16~\mu{\rm A\,m^{-1}}$ for FeSn/Py, which is an order of magnitude larger than the value $I_{\rm C}=3.9\pm 0.01~\mu{\rm A\,m^{-1}}$ measured for the standard Pt/Py bilayer system. This enhancement demonstrates that FeSn is a comparatively more efficient spin-charge converter than conventional heavy metal Pt. A large $I_{\rm C}$ and g$_{\rm eff}^{\uparrow\downarrow}$ in FeSn/Py film stack is attributed to its kagome-derived Berry curvature.~\cite{gupta2025symmetry}

\section{Conclusion}

In conclusion, we investigate the spin-dynamic properties of the epitaxial FeSn/Py bilayer heterostructure using FMR response. The results reveal a giant effective spin mixing conductance (g$^{\uparrow\downarrow}_{\rm eff})$  of $\approx (116\pm7)~{\rm nm}^{-2}$ for FeSn/Py, which is nearly one order of magnitude larger than that of conventional HM/FM systems. The substantial enhancement of g$^{\uparrow\downarrow}_{\rm eff}$ compared to reference Py films indicates highly efficient spin pumping at the FeSn/Py interface, underscoring the role of the kagome lattice in FeSn. The interfacial origin of the enhanced spin mixing conductance is further confirmed by an almost two-fold suppression of the effective damping upon insertion of a 3~nm Al spacer layer. As a result of giant g$^{\uparrow\downarrow}_{\rm eff}$, the FeSn/Py heterostructure also shows a nearly order-of-magnitude higher spin-to-charge conversion efficiency compared to the conventional Pt/Py stack. Together, these results establish FeSn as a promising kagome Dirac antiferromagnet for efficient spin–angular–momentum transfer and energy-efficient spintronic applications.

\section{Acknowledgments}
We gratefully acknowledge the partial support provided by the Ministry of Human Resource Development under the IMPRINT program (Grants Nos. 7519 and 7058), the Ministry of Electronics and Information Technology (MeitY) (Approval No. Y-20/8/2024-R\&D-E), the Science and Engineering Research Board (SERB, File Nos. CRG/2018/001012 and CRG/2022/002821), the Joint Advanced Technology Centre at IIT Delhi, the Grand Challenge Project at IIT Delhi, the Department of Science and Technology under the Nanomission program [Grant No. SR/NM/NT-1041/2016(G)], and the IEEE Magnetics Society Special Project [P.O. 502182]. The authors are thankful to Mr Sunil Ojha and Mr G. R. Umapathy, IUAC, New Delhi, for the RBS measurements. K.I.A.K. acknowledges support from the University Grants Commission (UGC), India. The authors acknowledge the support of the Central Research Facility, IIT Delhi, for providing characterization facilities.
\setcounter{figure}{0}  
\setcounter{table}{0}  
 \renewcommand{\thefigure}{S\arabic{figure}}
\renewcommand{\theequation}{S\arabic{equation}}
\renewcommand{\thetable}{S\arabic{table}}
\renewcommand{\thepage}{S\arabic{page}}

\renewcommand{\thesection}{S\arabic{section} :}
\setcounter{section}{0}
\renewcommand{\thesubsection}{S\arabic{subsection}}
\setcounter{subsection}{0}
\newpage
\section*{\Large \centering Supporting Information for: \\Topologically Driven Giant Effective Spin Mixing Conductance in Antiferromagnetic FeSn/Py Heterostructures}
\section{Dependence of linewidth and Resonance field.}
\begin{figure*} [htbp]
\centering
\includegraphics*[width=0.9\linewidth]{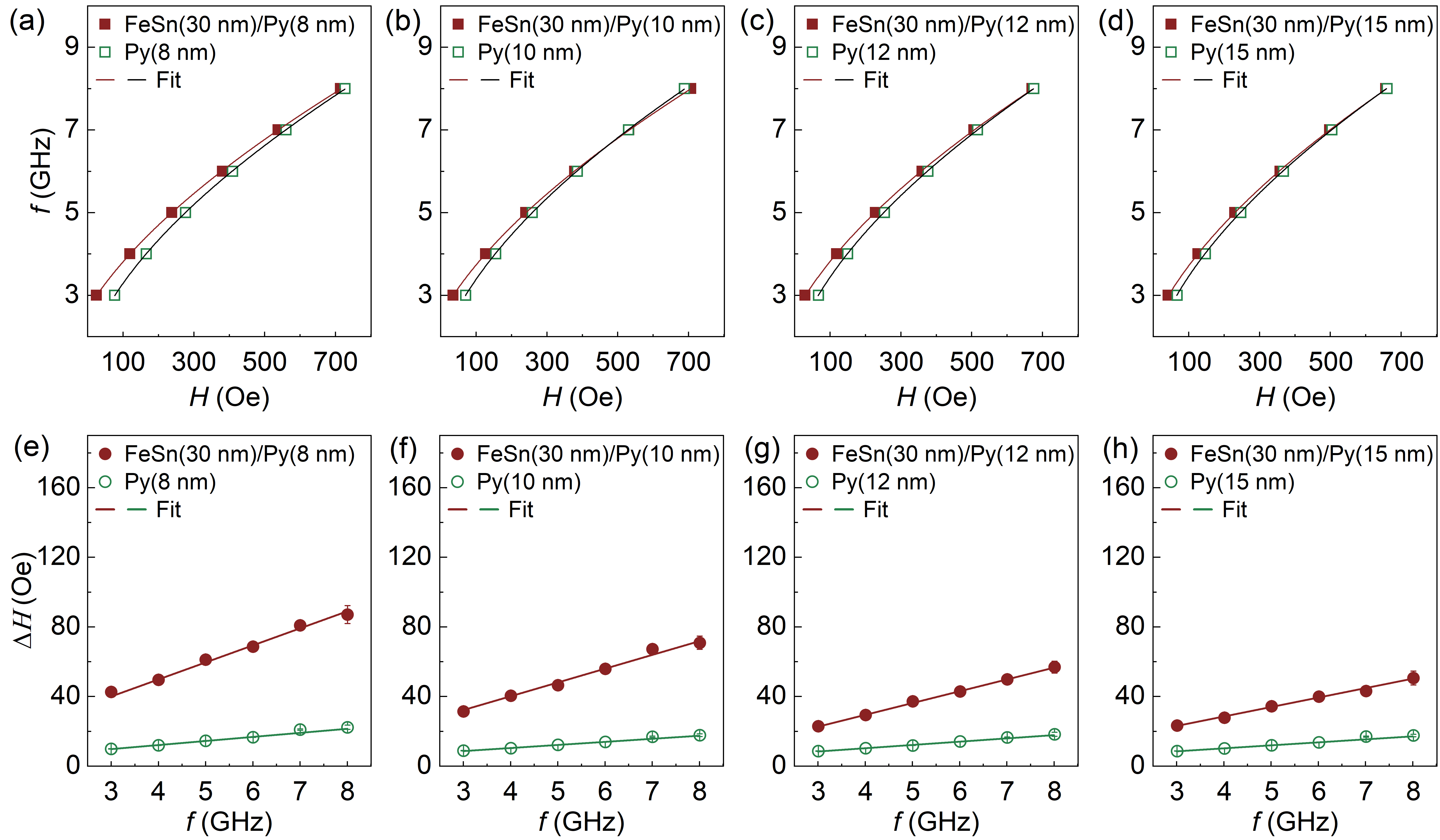}
\caption{\label{fig:RBS} (a-d) Dependence of the resonance frequency $f$ on the resonance field $H_{\mathrm{R}}$ with varying thicknesses (8 nm, 10 nm, 12 nm, 15 nm) of Py layer in FeSn/Py(5~nm) and ref: Py-stack. (e-h) Frequency dependence of the linewidth $\Delta H$ for the same set of samples for both stack. The symbols represent the experimental data, while the solid lines denote the fits.}
\end{figure*}
\providecommand{\noopsort}[1]{}\providecommand{\singleletter}[1]{#1}

\end{document}